# LinTO : Assistant vocal open-source respectueux des données personnelles pour les réunions d'entreprise

Jean-Pierre Lorré[1], Isabelle Ferrané[2], Francisco Madrigal[3], Michalis Vazirgiannis[4], Christophe Bourguignat[5]

[1]LINAGORA
[2]IRIT
[3]LAAS-CNRS
[4]Laboratoire d'Informatique de l'Ecole Polytechnique
[5]Zelros

jplorre@linagora.com, isabelle.ferrane@irit.fr, jfmadrig@laas.fr, mvazirg@lix.polytechnique.fr, christophe.bourguignat@zelros.com

## Résumé

*Cet article présente les premiers résultats du projet de recherche PIA Grands Défis du Numérique LinTO[1] dont l'objectif est de réaliser un assistant vocal permettant d'aider les employés d'une entreprise en particulier lors des réunions. Dispositif interactif doté de micros, écran et caméra 360°, il permet de piloter la salle, d'interroger le système d'information, d'aider à l'animation de la réunion et propose un environnement d'aide à la rédaction du compte rendu. Diffusé suivant un modèle ouvert respectueux des données personnelles, LinTO[2] est le premier assistant d'entreprise open-source conçu pour favoriser la prise en compte des exigences du RGPD.*

## Mots Clef

Assistant Vocal Conversationnel, Intelligence Artificielle, Reconnaissance Automatique de la Parole, Traitement du Langage Naturel, Reconnaissance de personnes

## Abstract

*This paper presents the first results of the PIA « Grands Défis du Numérique » research project LinTO. The goal of this project is to develop a conversational assistant to help the company's employees, particularly during meetings. LinTO is an interactive device equipped with microphones, a screen and a 360° camera, which allows to control the room, query company's information system, helps facilitate the meeting and provides an environment to aid minute writing. Distributed according to an open model that respects private data LinTO is the first open-source enterprise's assistant designed to comply with the GDPR requirements.*

## Keywords

Conversational Voice Assistant, Artificial Intelligence, Automatic Speech Recognition, Natural Language Processing, People Recognition.

## 1 Introduction

Nous assistons à une prolifération d'outils dans le domaine des assistants personnels. Présents sur les téléphones portables, ils aident à écrire des sms, des mails, prendre des rendez-vous ; associés aux outils de communication collaboratifs ils jouent le rôle d'un être humain connecté pour aider. Le marché associé à ces technologies est identifié par de nombreux analystes, certains l'évaluant à 5,1 milliards de dollars en 2022 [5] en croissance annuelle de 32% entre 2015 et 2022.

LinTO est un assistant conversationnel vocal offrant des fonctionnalités avancées et adaptées au milieu professionnel. À la différence des assistants personnels grand public, LinTO est conçu pour s'interfacer avec une plateforme de productivité (telle que OpenPaaS, Office 365 ou G Suite), les autres briques du système d'information de l'entreprise ainsi que des services externes. Il dispose ainsi de fonctions propres à l'assistance dans un contexte de travail : gestion de réunion, compte-rendu, accès à des données pour la prise de décision, etc.

Remarquons que toutes les entreprises des "GAFAM" (Google, Amazon, Facebook, Apple et Microsoft) proposent une offre d'assistant intelligent. Les acteurs désirant se positionner sur ce marché se retrouvent souvent face à un choix cornélien: ou bien entrer en compétition R&D avec ces acteurs possédant des moyens bien supérieurs, ou bien utiliser et dépendre des services de leurs concurrents directs tout en les enrichissant de nouvelles données. Une autre voie est donc nécessaire. Le modèle open-source a démontré dans d'autres domaines qu'il était possible à une communauté de "petits" acteurs collaborant et mutualisant leurs connaissances de mettre à mal la situation hégémonique d'un acteur dominant un marché. L'objectif de ce papier, est de présenter les premiers résultats du projet collaboratif LinTO soutenu par le Programme des Investissements d'Avenir.

## 2 Assistant LinTO

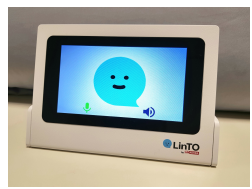

Le dispositif matériel "LinTO" inclut carte CPU de type Raspberry Pi, écran tactile, haut-parleurs, matrice de microphones et suivant les configurations, une caméra 360°. Il est complété par une plateforme logicielle d'assistant conversationnel supportant différentes modalités d'interaction en fonction des besoins et des configurations matérielles.

Deux groupes de fonctionnalités complémentaires sont étudiés :

- un ensemble de fonctionnalités de type "assistant personnel" pour aider l'utilisateur à accéder à l'information qu'elle soit interne au système

---

d'information (mail, rendez-vous, document, etc.) ou externe par l'intermédiaire de la connexion avec un service disponible sur Internet ;
- un second ensemble de fonctionnalités dédiées au contexte de la réunion, incluant des mécanismes d'aide à la modération de réunion, de reconnaissance de participants et de suivi de leurs échanges, de recommandations contextuelles et de génération semi-automatique de résumé. C'est dans ce groupe que sont situés les principaux verrous du projet, en particulier ceux liés à la détection de plusieurs personnes (contexte multi-participants), à la définition d'une signature audio-visuelle (caractériser sans chercher à identifier) et à la fusion d'indicateurs issus d'analyses bas niveau (sonore/visuel) [2] pour enrichir les traitements sur l'analyse des interactions.

Compte tenu de l'usage dans des entreprises disposant d'un nombre élevé de salles de réunions, un outil d'administration est également proposé afin de gérer à distance la flotte des LinTO.

Le contexte professionnel induit des contraintes. Outre la problématique de la sécurisation des données confidentielles, l'accès aux informations pertinentes est souvent conditionné par l'intégration avec les systèmes et processus existants déjà mis en place au sein de l'entreprise. Enfin, dans un contexte professionnel les informations utiles à l'assistant ne sont pas nécessairement regroupées autour d'un seul individu, mais peuvent nécessiter de croiser celles de plusieurs participants (par exemple la sélection de date pour une réunion nécessite l'accès aux calendriers des différents participants). Pour obtenir une expérience utilisateur correcte dans un tel cadre, il faut donc être capable d'offrir des performances satisfaisantes en termes de reconnaissance de la parole et de traitement du langage naturel (justesse de la transcription, temps de réponse et montée en charge), mais il est également important de pouvoir s'interfacer avec une plateforme d'entreprise capable de fournir les informations nécessaires à ces fonctionnalités avancées.

## 3 Reconnaissance de la parole en réunion

Malgré les grandes avancées dans le domaine, la reconnaissance de la parole reste un défi technique important [3]. Bien que cette dernière soit désormais suffisamment bien maîtrisée pour donner des résultats satisfaisants suivant le type d'usage (dialogue de commande ou dictée vocale), son application au dialogue entre humains dans le cadre d'une réunion reste un verrou. Notons qu'il n'existe pas de solution open-source disponible pour le français.

LINAGORA développe un moteur de reconnaissance automatique de la parole (ASR, Automatic Speech Recognition) nommé LinSTT. Ce dernier est basé sur la boîte à outils open-source Kaldi [9].

Les deux types d'usages proposés précédemment reposent sur deux modes de fonctionnement du composant ASR : (i) mode "commande", il s'agit alors de reconnaître une commande et ses attributs dans un contexte bien défini ; (ii) mode "large vocabulaire" où le dispositif doit être à même de transcrire un flux de parole concernant des sujets ouverts.

## 4 Mode commande

La chaîne de traitements mise en œuvre par la plate-forme LinTO en mode commande est illustrée ci-dessous. L'outil traite en permanence le son provenant d'un ou plusieurs microphones, se déclenche à l'énoncé du « hotword », détecte l'activité vocale, repère l'énoncé (utterance), extrait de la transcription les principaux concepts (intention, entités) qui permettent de constituer l'action à exécuter (skill).

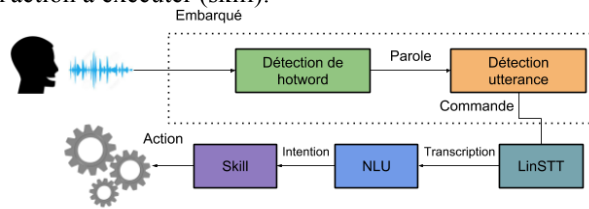

Figure 1 : Chaîne de traitement en mode commande.

### 4.1 Détection du « hotword »

Le mode commande est déclenché à l'aide d'un mécanisme de réveil qui consiste à reconnaître un ou plusieurs mots spécifiques prononcés par l'utilisateur afin de s'activer.

Dans cette approche la détection d'un mot-clef (*hotword*) prononcé par l'utilisateur précède une commande ou une question. Le dispositif sait que la phrase suivant le hotword lui est destinée et il ne traite que celle-ci. La reconnaissance du mot de réveil se base sur un classifieur neuronal de type réseaux récurrents. Les LSTM (Long Short Term Memory) et GRU (Gated Recurrent Units) sont alors deux solutions possibles. Le réseau GRU est une généralisation d'un réseau LSTM et offre de meilleurs résultats sur des jeux de données restreints, de plus, à performance égale il est plus léger qu'un réseau LSTM [12]. C'est donc ce type de réseau qui a été retenu.

Quand le mot-clef est détecté, le signal est analysé afin de détecter l'activité vocale pour identifier le début et la fin d'une commande. Cette commande est ensuite envoyée à LinSTT qui transcrit l'audio en texte à partir duquel est extraite une intention. Cette dernière est associée à une compétence matérialisée par un composant logiciel (un *skill*) qui effectue une action comme par exemple projeter un document.

### 4.2 Détection d'intention

L'analyse du texte transcrit permet d'identifier les intentions et leurs paramètres en vue d'activer les actions associées (exemple : LinTO *allume* la lumière).

Pour cette étape nous nous appuyons sur le composant open-source TOCK [15] qui permet de construire des modèles d'analyse du langage naturel. Ce dernier propose une interface graphique qui permet d'associer à chaque phrase l'intention et les entités correspondantes afin de construire des modèles. Cette étape utilise les librairies Stanford CoreNLP ou Apache OpenNLP.

Nous obtenons des performances respectables, tant dans le cas "commande" (taux de reconnaissance des intentions supérieur à 97%) que dans le cas large vocabulaire (Word Error Rate, WER inférieur à 14% évalué sur le corpus ESTER [4], modèle DNN-HMM).

## 5 Fonctionnalités avancées en réunion

Il s'agit ici d'analyser la conversation afin de proposer l'aide à la décision, l'extraction de thèmes en temps réel et le résumé automatique.

## 5.1 Interaction langagière multi-locuteurs

Les services d'aide à la gestion de réunions ont été identifiés dans [11] comme des scénarios complexes, centrés sur l'interaction conversationnelle. La multimodalité en situation multipartite a été étudiée dans et représente toujours un verrou important [2].

A partir des transcriptions fournies par les étapes précédentes, le projet prévoit une analyse des interactions, pour répondre à trois questions : (1) à qui s'adresse-t-on dans le cas d'un dialogue à plus de deux personnes (2) quels sont les buts conversationnels d'une intervention (réponse à une question, accusé de réception, élaboration d'une question passée, acceptation d'une proposition, etc.), ce que l'on désigne par le terme d'actes de dialogue et (3) quels sont les fils conversationnels qui composent l'interaction à plusieurs. Pour pallier le coût de constituer des données annotées utilisées dans la plupart des approches automatiques qui reposent sur de l'apprentissage supervisé, le projet a pour but de développer de la supervision distante, en suivant l'approche décrite dans [10]. Ces travaux sont en cours et font l'objet du focus particulier sur le deuxième année.

## 5.2 Synthèse de réunion

Afin de produire un résumé à la fin de la réunion, deux approches seront comparées. La première vise à produire d'un seul jet un résumé pouvant être lu et sauvegardé tel quel, en se passant de toute intervention humaine. La seconde approche, produit une proposition de résumé sous la forme d'un modèle pré-rempli, demandant ensuite à être corrigé et réorganisé par un ou plusieurs participants. Lors de la première année du projet, c'est la première approche qui a été évaluée.

Nous avons travaillé sur la compréhension par la machine du texte issu du système ASR. Cette première étape indispensable permet ensuite de grouper les utterances de telle sorte que pour chaque groupe, une phrase résumant l'ensemble du groupe puisse être produite. Le compte rendu final est alors composé de l'ensemble de ces phrases résumées.

Plus précisément, un encodeur neuronal d'utterances basé sur des réseaux récurrents et trois types de mécanismes attentionnels [1] a été développé. Cet encodeur a ensuite été incorporé dans des architectures siamoises et triplettes [8] et entraîné sur le corpus AMI [6]. Les évaluations montrent que l'encodeur proposé permet d'obtenir des groupes d'utterances proches des solutions humaines, et généralement meilleurs que ceux retournés par d'autres systèmes de l'état de l'art [14].

Par ailleurs, le corpus AMI est en cours de traduction via une approche de *crowdsourcing* afin de pouvoir développer une version française du modèle.

## 6 Reconnaissance visuelle

LinTO doit inclure des services de reconnaissance de participants et de gestes à l'aide d'une caméra 360° pour apporter une assistance supplémentaire. Le traitement audio est incapable seul de répondre à une partie des besoins comme le comptage lors d'un vote collectif ou la demande d'une prise de parole en réunion.

La reconnaissance visuelle vise également à inférer une localisation topologique des participants à la réunion afin de permettre à LinTO d'agir sur la matrice de microphones et renforcer la reconnaissance audio.

De manière globale, deux stratégies de fusion des percepts audio et visuel sont à considérer pour notre assistant: d'un coté, l'information provenant du module vision est récupérée puis synthétisée vocalement par LinTO comme par exemple lors d'une demande de parole à distance. De l'autre côté, c'est un participant qui demande à LinTO d'exécuter une tâche nécessitant la ressource vision par exemple le comptage des personnes « pour » dans le cas d'un vote.

Cependant, l'intégration de la reconnaissance visuelle dans LinTO pose plusieurs défis :
- traitement du flux vidéo d'images 360° ;
- construction d'un corpus d'images pour des personnes en réunion, annotation et entraînement de modèles d'apprentissage profond ;
- contrainte temps réels.

Notons que nous nous focalisons sur la localisation des participants à la réunion et pas à l'identité des individus.

### 6.1 Dispositif dédié à l'analyse visuelle

Notre dispositif privilégie une caméra RICOH THETA pour son faible coût, sa popularité auprès des développeurs et la possibilité de streaming vidéo. Les images sont acquises au format «dual-ficheyes» puis transformées en images panoramiques prêtes pour le traitement. Il est judicieux d'étudier les modèles d'apprentissage profond existants afin d'évaluer leurs performances dans notre contexte applicatif. De plus, ces modèles sont déjà entraînés et feront gagner un temps considérable quand à la construction de corpus, l'annotation et l'entraînement.

### 6.2 Reconnaissance de participants

L'objectif est de détecter les personnes en réunions et les repérer spatialement en temps réel à partir de notre caméra. Comme évoqué, la localisation des personnes en environnement encombré donc en présence d'occultations éventuelles (écran, autres personnes, etc.), est un verrou important.

Notre détecteur de personnes est basé sur le modèle SSD Mobilenet (mieux adapté à des cartes matérielles de faible ressources) et un filtrage spatio-temporel pour le suivi (*tracking*). Le principe est le suivant :
- le filtrage spatial repose sur le recouvrement entre les régions images associées aux détections de personnes entre deux instants d'image ;
- le filtrage temporel repose sur l'hypothèse de faible déplacement des personnes entre images successives. On estime cette valeur à 6 secondes après des expérimentations pour une personne qui quitte sa place.

Ce filtrage est possible grâce à l'historique des détections mémorisé durant le déroulement de la réunion. L'ajout de l'algorithme DELF (*DEep Local Feature*) de *matchings* d'images permet de lever les doutes quand il y a confusion. Les problèmes qui persistent sont les non-détections sporadiques et les ambiguïtés lors d'un chevauchement de zones de détection durant le déplacement de personnes.

### 6.3 Reconnaissance de gestes

Le modèle présenté dans [7], entraîné sur un million d'images de gestes, a été expérimenté avec succès sur nos images et semble pertinent pour la reconnaissance de gestes. Cependant nos investigations actuelles ne permettent pas de cibler à ce jour le pipeline adapté à l'intégration de nos modalités compatibles avec les ressources matérielles limitées.

### 6.4 Discussion sur l'analyse visuelles

Ces évaluations qualitatives préliminaires ont permis d'exhiber quelques verrous listés ci-après :

- mouvements conjoints : expressions faciales, rotations extrêmes de la tête, etc. induisant des erreurs d'interprétation, par exemple, une personne qui se tourne pour parler à la personne d'à côté ;
- variation d'éclairage de la salle de réunion induisant des problèmes de robustesse, par exemple, des salles avec des fenêtres donnant sur l'extérieur ou dans des situations où la lumière s'éteint pendant une présentation et s'illumine pendant la discussion ;
- distance importante caméra / scène induisant des résolutions images trop faibles pour inférer certains percepts comme des gestes faciaux, cela se produit généralement lorsque on place la caméra sur un mur ou un coin de la salle de réunion.
- variabilité du point de vue caméra / scène induisant des problèmes de robustesse, les participants sont occultés par la personne la plus proche de la caméra.

La question de la performance demeure centrale, notamment pour un couplage de traitement audio/vidéo. Plusieurs solutions sont en cours d'expérimentation comme par exemple la délégation de certains traitements à distance.

## 7 Respect des données personnelles

Un facteur important du projet concerne la sécurité et la confidentialité des données personnelles. En effet, l'objectif de LinTO est d'aider les participants présents en salle de réunion. Afin de pouvoir les suivre efficacement, LinTO est susceptible d'écouter leurs conversations et d'analyser diverses données les concernant (emails, calendriers, etc.). Afin de minimiser les risques d'atteinte à la vie privée, le traitement doit être effectué en local (dispositif LinTO dédié).

Pour cela LinTO traite toutes les données à l'intérieur du système d'information de l'entreprise, éliminant ainsi le besoin d'envoyer les données vers un Cloud externe et ne stocke pas les données vocales. Par conséquent, personne ne dispose d'informations sur les paroles échangées, ce qui protège contre le piratage et la surveillance de masse.

## 8 Prochaines étapes

Le projet a commencé en avril 2018 et doit se terminer en mars 2021. La première année a permis de mettre en place le projet et de valider le mode assistant personnel qui s'appuie sur le modèle commande. Des prototypes sont disponibles pour l'aide à la rédaction de compte-rendu de réunions [13] ainsi que pour la localisation spatiale des participants. Les prochaines étapes concernent l'étude des stratégies de fusion des percepts audio et visuel, l'intégration des approche d'analyse des interactions langagière avec la démarche de génération de résumé et enfin la qualification définitive de l'architecture du dispositif LinTO afin de prendre en compte les contraintes aussi bien d'un point de vue scientifique que du traitement des données personnelles.

## Bibliographie